\documentstyle[epsf, rotate]{article}
\begin{document}

\begin{center}
{\Large
        Magnetic fields of neutron stars in X-ray pulsars
}
\vskip 0.5cm
{\bf   V.M. Lipunov$^{1,2}$ \& S.B. Popov$^1$}\\

$^1$Sternberg Astronomical Institute\\
$^2$Department of Physics, Moscow State University\\

\vskip 0.5cm

119899, Russia, Moscow,\\
Universitetski pr. 13\\
e-mail: lipunov@sai.msu.su; polar@xray.sai.msu.su\\

\vskip 0.5cm

{\bf    Abstract}

\end{center}

 Estimates of the magnetic field of neutron stars in X-ray pulsars
are obtained using the hypothesis of the equilibrium period for disk and
wind accretion and also
from the BATSE data on timing of X-ray pulsars
using the observed maximum spin-down rate.
Cyclotron lines at energies $\ge 100$ keV in several Be-transient are
predicted for future observations.\\


%\newpage

\section{Introduction}

Among all astrophysical objects neutron stars (NSs)
attract most attention of physicists.
Now we know more than 1000 NSs as 
radiopulsars and more than 100 NSs emiting X-rays,
but the Galactic population of these objects is about $10^8$ -- $10^9$.
 Here the first number comes mainly from radiopulsars statistics, 
and should be
considered as a low limit, because it is not clear if all NSs pass through
the stage of a radiopulsar, as far as initial parameters (spin period and
magnetic field) of significant part of NSs can be different from ``standard''
values: $B\sim10^{12}$ G, $p\sim 1-20 $ ms. For example, NSs can be born below
the death-line due to small initial magnetic fields, or relatively long 
periods (fall-back after a supernova explosion also can be important,
because magnetic momentum or spin period can be changed in that process). 
And the second number is in
correspondence with models of chemical evolution of the Galaxy.
So only a tiny fraction of one of the most fascinating astrophysical objects
is observed at present. 

 NSs can appear as sources of different nature:
as isolated objects (radio pulsars, 
old isolated accreting NSs, soft $\gamma$-- repeaters etc.) 
and as binary companions, usually as 
X-ray sources in close binary systems, powered by wind or disk
accretion from a secondary companion.
X-ray pulsars are probably one of 
the most prominent among these sources, because there important
parameters of NSs (spin period, magnetic field etc.) can be determined.

 Now we know more than 40 X-ray pulsars 
(see e.g. Bildsten et al. 1997, Borkus 1998). Observations of optical
counterparts of X-ray sources
give an opportunity to determine distances to these objects 
and other parameters with relatively
high precision, and with hyroline detections one can obtain the value of
magnetic field, $B$, 
of a NS. But lines are not detected in all sources of that
type (partly because they can lay out of the range of necessary
spectral sensitivity of devices, when fields are too high,
$>10^{13}$ G, for example), 
and magnetic field can be estimated
from period measurements (see e.g. Lipunov 1982, 1992). Precise distance
measurements usually are not available immediately after X-ray discovery
(especially, if localization
error boxes are large and X-ray sources have transient nature).
In that sense methods of simultaneous determination of field and distance
basing only on X-ray observations can be useful, and several of them were
suggested by different authors previously.

 Here we try to obtain estimates of the magnetic fields (and distances)
of NSs in X-ray pulsars from their period (and flux) variations.

%\newpage 

\section{Estimates of the magnetic field}

 Magnetic fields of accreting NSs can be estimated using period
variations or using the hypothesis of the equilibrium period (see Lipunov
1992). We use both of these methods.

 For estimating of magnetic momentum of NSs using observed values
of maximum spin-down we use the following main equation:

$$
\frac{dI\omega}{dt}=-k_t\frac{\mu^2}{R_{co}^3},
$$
where $I$ -- NS's momentum of inertia, $\omega=\frac{2\pi}p$ 
-- spin frequency, $\mu$ -- magnetic momentum, $R_{co}=\left
(\frac{GM}{\omega^2}\right)^{1/3}$-- corotation radius. 
We used $k_t=1/3$, $I=10^{45}$ g cm$^2$, $M=1.4M_{\odot}$.
We can use this approximation with no spin-up )accelerating) momentum,
because we choose moments with maximum spin-down, when spin-doqn
(braking) momentum is much larger than accelerating momentum.
This estimate normally should be considered as a low limit on the value of  
the magnetic field.

We used graphs from (Bildsten et al., 1997) to derive spin-up
and spin-down rates and flux changes measurements.  Data on these graphs is  
shown with one day time resolution. Usually errors are relatively small, and
we neglecte them. 

 Such estimates were obtained several times by different authors
with different data sets, but usually these sets had had worse time
resolution
(see some examples in (Lipunov 1992)). And the BATSE data (Bildsten et al.,
1997) gives
an excellent opportunity to repeat these simple calculations.

 Equilibrium period can be written in different forms for disk and wind-fed
systems. For the first case we used the following equation:

\begin{equation}
p_{eq. disk}=2.7\, \mu_{30}^{6/7} L_{37}^{-3/7}\, s.
\end{equation}

For wind-accreting systems we have:

\begin{equation}
p_{eq. wind}=10.4\,L_{37}^{-1}T_{10}^{-1/6}\mu_{30} \, s.
\end{equation}
Here $L_{37}$ -- luminosity in units $10^{37}$ erg s$^{-1}$,
$T_{10}$ -- orbital period in units 10 days, $\mu_{30}$ -- magnetic momentum
in units $10^{30}$ G cm$^{3}$. 

 Estimates of the magnetic momentum, $\mu$, obtained with different
assumptions are shown in the table 1.
Three values are shown: an estimate from spin-down obtained from the BATSE
data (Bildsten et al., 1997); 
an estimate from the equilibrium period for wind-fed systems (eq. (2));
an estimate for disk-accreting systems (eq. (1)). 
Both of the last two estimates
were made for X-ray pulsars about which we were not sure if they are
disk or wind-accreting systems, less probable values 
(wind accretion in Be-transients) are marked with asterix. 

\begin{table}[h]
\caption[]{Spin-down and magnetic momentum estimates}
\begin{tabular}{|l||c|c|c|c|c|}
\hline
X-RAY & maximum & Source & Magnetic & Magnetic & Magnetic \\
 PULSAR & dp/dt   & Type & momentum & momentum &momentum \\
        & observ. &     & (spin-down), & (wind), &(disk),\\
        & (spin-down)           &     & $10^{30}$ G cm$^{3}$ &$10^{30}$ G cm$^{3}$ & 
$10^{30}$ G cm$^{3}$ \\
\hline
 GRO 1744-28    &         &   BeTR   &         &    0.93$^*$  &  0.58\\
 HER X-1     & $9.3\cdot10^{-13}$ &   LMXRB  &     0.3 &          &  0.18\\
 4U 0115+63  & $3.0\cdot10^{-10}$ &   BeTR   &    5.17 &    0.32$^*$  &  1.26\\
 CEN X-3     & $7.5\cdot10^{-12}$ &   HMSG   &   0.82  &    1.8   &  4.42\\
 4U 1627-67  & $4.1\cdot10^{-11}$ &   LMXBR  &   1.9   &          &  2.82\\
 2S 1417-624 &         &   BeTR     &         &    8.64$^*$  &  17.82\\
 GRO 1948+32 & $5.4\cdot10^{-9}$  &   BeTR   &   22.0  &          &      \\
 OAO 1657-415& $1.5\cdot10^{-7}$  &   HMSG   &   115.1 &    0.15  &  4.33\\
 EXO 2030+375&         &   BeTR   &         &    0.1   &  3.45\\
 GRO 1008-57 & $3.2\cdot10^{-8}$  &   BeTR   &   53.3  &          &      \\
 A 0535+26   &         &   BeTR   &         &    30.24$^*$ &  101.23\\
 GX 1+4      & $6.4\cdot10^{-8}$  &   LMXRB  &   75.5  &          & 167.3\\
 VELA X-1    & $3.8\cdot10^{-9}$  &   HMSG   &   18.5  &    4.03  &  88.15\\
 4U 1145-61  & $3.3\cdot10^{-7}$  &   BeTR   &   172.1 &    0.23$^*$  &  16.7\\
 A 1118-616  & $5.1\cdot10^{-7}$  &   BeTR   &   212.8 &          &  245.5\\
 4U 1535-52  & $3.6\cdot10^{-7}$  &   HMSG   &   56.4  &    17.37 &  299.3\\
 GX 301-2    & $8.9\cdot10^{-6}$  &   HMSG   &   281.9 &    8.34  &  200.5\\
\hline
\end{tabular}
\end{table}

In the table 2 we show values, which
were used for estimates with the hypothesis of
the equilibrium period: spin period,
mean luminosity in units $10^{37}$ erg s$^{-1}$, orbital period in units 10
days (see a compilative catalog of X-ray pulsars in the Web at the URL:\\
{\bf http://xray.sai.msu.ru/\~\,polar/html/publications/cat/x-ray\_n2.www}). 
In table 1 we use the following
notation: LMXRB- Low Mass X-Ray Binary;
HMSG - High Mass SuperGiant;
BeTR- Be-transient source.

\begin{table}[h]
\caption[]{Parameters of X-ray pulsars}
\begin{tabular}{|l||c|c|c|}
\hline
  X-RAY &  Period, &  Mean & Orbital\\
 PULSAR &  sec     &  luminosity, & Period,\\
        &          &  $10^{37} \,{\rm erg}\, {\rm s}^{-1}$& days\\
\hline
GRO 1744-28     &      0.467  &      20      &       11.76\\
HER X-1      &      1.24  &        0.2    &            \\
4U 0115+63   &      3.61  &        0.8    &       24.3\\ 
CEN X-3      &      4.84  &        5   &            \\
4U 1627-67   &      7.66  &        0.7 &           0.0289\\
2S 1417-624  &     17.6  &         4   &          42.1   \\
GRO J1948+32 &     18.7   &            &                 \\
OAO 1657-415 &     37.7   &        0.04 &         10.44 \\ 
EXO 2030+375 &     41.7    &       0.02    &      46.0\\   
GRO 1008-57  &     93.5     &      & \\
A 0535+26    &    105      &       2      &       110\\
GX 1+4       &    120      &       4       &        \\
VELA X-1     &    283      &       0.15     &      8.96\\
4U 1145-61   &    292      &       0.005    &    187   \\
A 1118-616   &    406.4    &       0.5      &          \\
4U 1535-52   &    530      &       0.4      &      3.73\\
GX 301-2     &    681     &        0.1      &     41.5\\ 
\hline
\end{tabular}
\end{table}

 More precise estimates can be made by fitting all observed values
of spin-up and spin-down rate together with flux measurements. 
When the distance to the source is know
only the value of the magnetic field should be fitted.
And on figures 1-2 we show such estimates for two X-ray pulsars.

\begin{figure}
\epsfxsize=0.9\hsize
\centerline{\rotate[r]{\epsfbox{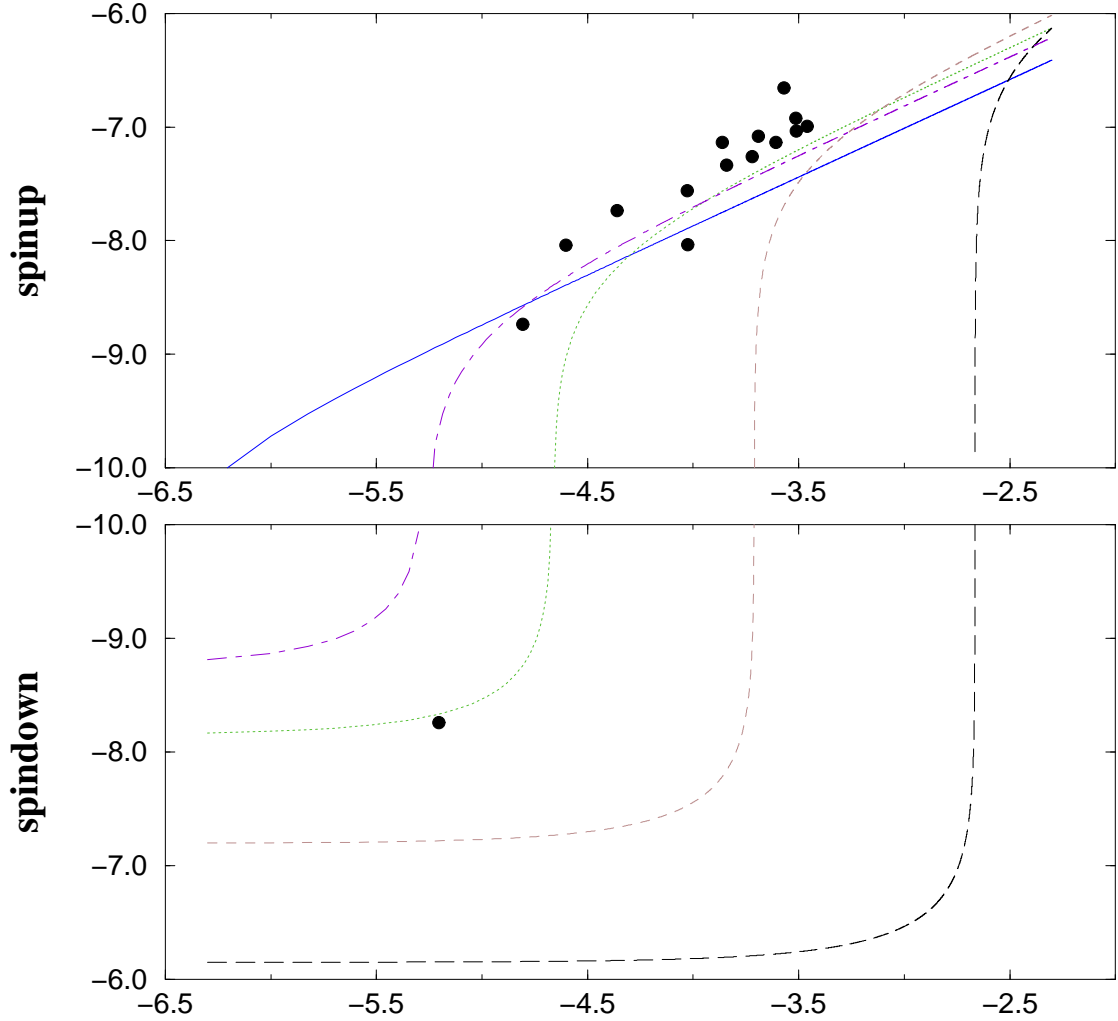}}}
\caption{Dependence of period derivative, $\dot p $, on the parameter $p^{7/3}f$,
$f$-- observed flux, for A0535+26. Both axis are in logarithmic scale.
Observations (Bildsten et al., 1997) are shown with black dots.
Five curves are plotted for different values of the magnetic field.
 Solid curve: $\mu=1\cdot 10^{30}\, {\rm G}\cdot {\rm cm}^3$.
Dot-dashed curve: $\mu=5\cdot 10^{30}\, {\rm G}\cdot {\rm cm}^3$.
Dotted curve: $\mu=10\cdot 10^{30}\, {\rm G}\cdot {\rm cm}^3$.
Dashed curve: $\mu=30\cdot 10^{30}\, {\rm G}\cdot {\rm cm}^3$.
Long dashed curve: $\mu=100\cdot 10^{30}\, {\rm G}\cdot {\rm cm}^3$.
All curves are plotted for the distance $d=2 \, {\rm kpc}$.}
\end{figure}

\begin{figure}
\epsfxsize=0.9\hsize
\centerline{\rotate[r]{\epsfbox{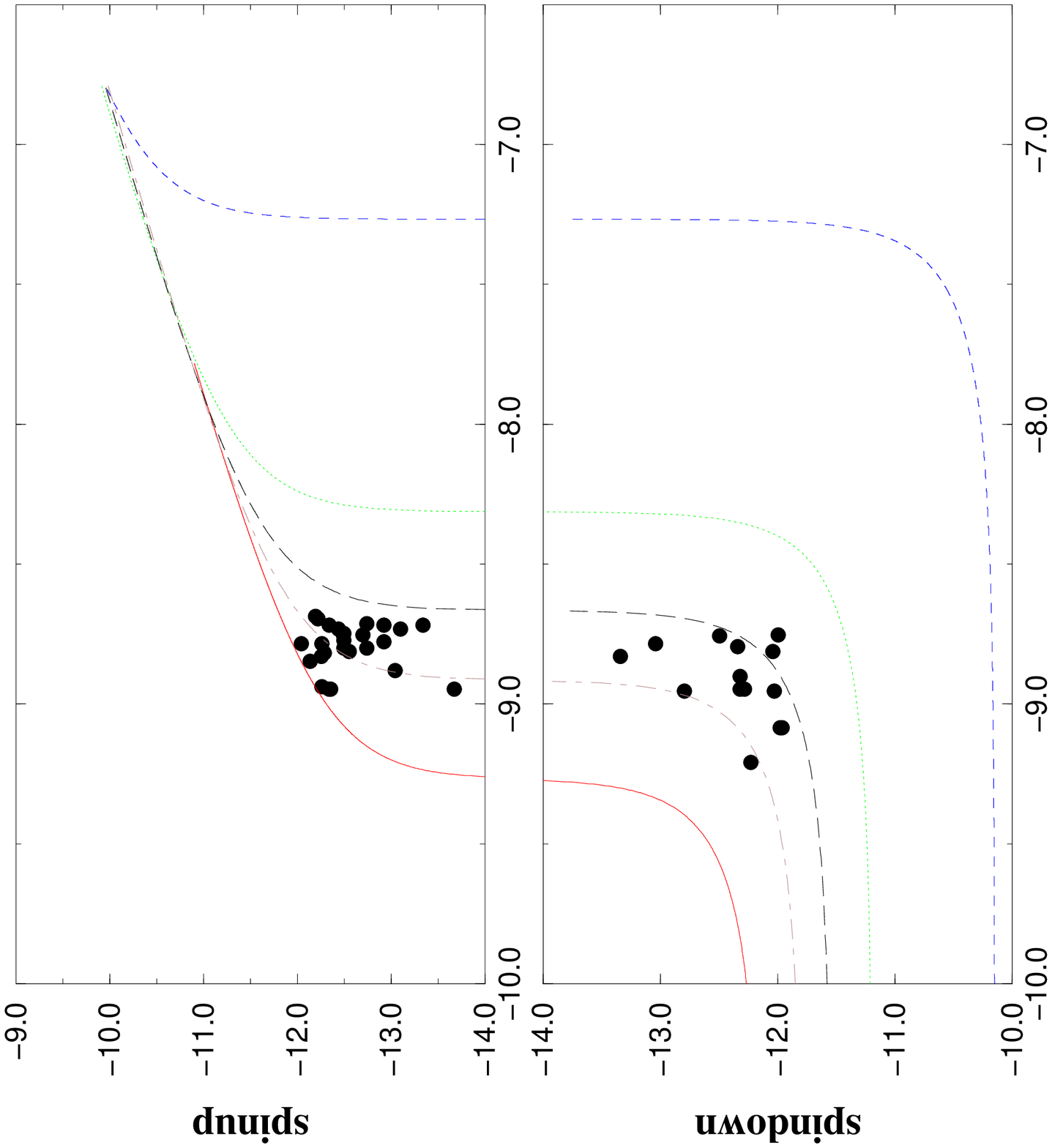}}}
\caption{Dependence of period derivative, $\dot p $, on the parameter $p^{7/3}f$,
$f$-- observed flux, for Her X-1. Both axis are in logarithmic scale.
Observations (Bildsten et al., 1997) are shown with black dots.
Five curves are plotted for different values of the magnetic field.
 Solid curve: $\mu=0.1\cdot 10^{30}\, {\rm G}\cdot {\rm cm}^3$.
Dot-dashed curve: $\mu=0.15\cdot 10^{30}\, {\rm G}\cdot {\rm cm}^3$.
Long dashed curve: $\mu=0.2\cdot 10^{30}\, {\rm G}\cdot {\rm cm}^3$.
Dotted curve: $\mu=0.3\cdot 10^{30}\, {\rm G}\cdot {\rm cm}^3$.
Dashed curve: $\mu=1\cdot 10^{30}\, {\rm G}\cdot {\rm cm}^3$.
All curves are plotted for the distance $d=4 \, {\rm kpc}$.}
\end{figure}

 We plot spin-up and spin-down rates as a function of the parameter,
which is a combination of the spin period and source's luminosity.
Spin-up and spin-down values derived
from the BATSE data (Bildsten et al., 1997)
are plotted as black dots,
and theoretical curves for different values of the magnetic momentum
are also shown. In ideal, the best curve for the magnetic momentum should
exist, which fits all observational points. In reality points have some
errors, distance to the source in also know with some uncertainty, and
simple model of spin-up and spin-down can be only the first approximation.
But these estimates of the magnetic momentum are more precise, than the
ones obtained with the equilibrium hypothesis.

 These 
estimates can be different from other ones obtained from the equilibrium
periods or from a single value of spin-down as can be seen from the table 1.

%\newpage

\section{Discussion and conclusions}

 We made estimates of the magnetic field of NSs in X-ray pulsars.
Estimates which were 
made with an assumption that $p=p_{eq}$ are rather rough.
Obtained values depend (except uncertainties connected with the method
itself) on unknown parameters of NSs, such as masses, radii, moments
of inertia. All of them were accepted to have ``standard'' values,
and of course it is only the first approximation.
For example, our estimate for the source GRO 1744-28 is $\mu \sim 10^{30}$
G cm$^3$, and it is smaller than the estimate shown in (Borkus 1998),
which is $B\sim (2-5)\cdot 10^{12}$ G (we mark, that the estimate obtained by
Joss \& Rappaport (1997) is significantly
lower than both: Borkus and our estimates). 
But if one take ``non-standard'' value for
$R$, these estimates of $\mu$ and $B$ can be in good correspondence.

We show several examples in table 3. NSs radii are calculated from the
following simple formula:

$$
R=\left(2\mu/B\right)^{1/3}.
$$
Here $\mu$ are taken from table 1, and values of $B$ are taken from Nagase
(1992), Borkus (1998) and Wang (1996). As one can see from the table
for several sources
measured $B$ are not in correspondence with our calculated $\mu$,
and radii of NSs are too big.
Mostly these cases are long period wind-fed pulsars like GX 301-2, 
where formation of temporal
reverse disk is possible for the cases of fast spin-down, 
so there maximum spin-down can be not the best field
estimate, and estimates from the equilibrium period for wind-accretion case 
are in better correspondence with observations. 
For A 0535+26 our estimate was obtained only from the equilibrium period.
And as far as this system is transient it can be far from the equilibrium.
We note, that in general existence of high magnetic field in that source,
as it comes from our estimates, is confirmed by observations.
In the case of 4U 0115+63 errors for maximum spin-down rate are significant,
and discrepancy between observed and calculated values can be due to this.
We also note, that Ginga was not sensitive enough
at the spectral region $\ge 40$ keV, where cyclotron line for $\mu \ge
(2-3)\cdot 10^{30}$ G cm$^3$ are situated.

\begin{table}[h]
\caption[]{Magnetic fields, magnetic momentum and radii}
\begin{tabular}{|l||c|c|c|}
\hline
X-RAY   & Magnetic &    Magnetic &   Neutron \\
 PULSAR & momentum &    field    &   star    \\
        & (calcul.), &  (observ.), &   radius,  \\
        & $10^{30}$G cm$^{3}$ & $10^{12}$G  & km \\
\hline
 GRO 1744-28    & 0.58 & $ \sim (2-5)$& $\sim (8.3-6.1)$ \\
 HER X-1     & 0.3  & 3& 5.8 \\
 4U 0115+63  & 5.17 & 1.1& 21.1 \\
 A 0535+26   & 101.23 &11& 26.4 \\
 VELA X-1    & 18.5 & 2.3 & 25.2 \\
 4U 1535-52  & 56.4 & 1.9 & 39 \\
 GX 301-2    & 281.9 & 3.5 & 54.4 \\
\hline
\end{tabular}
\end{table}

In more clear cases (Her X-1, GRO 1744-28), where we are sure, that
accretion is of the disk type, our estimates from maximum spin-down are in
good correspondence with observations. And we predict for the cases
of Be-transients, where disk accretion is working for sure, 
that in 2S 1417-624,
GRO 1948+32, GRO 1008-57, A 1118-616 and 4U 1145-61 
observations of cyclotron lines at
energies $\ge 100$ keV are possible in future.

 Estimates obtained from maximum spin-down rate and estimates obtained
with the hypothesis of equilibrium period are in rough correspondence,
except sources OAO 1657-415 and 4U 1145-61, where maximum 
spin-down estimates are
significantly higher. It can be an indication, that systems are far from
equilibrium (especially in the case of Be-transient  4U 1145-61), 
or that some additional mechanism of spin-down (outflows, reverse disks, ...?)
work.
In the case of OAO 1657-415 estimate based on maximum spin-down rate 
can be incorrect similar to GX 301-2 due to the reasons, which were
discussed above.

 Observations of period and flux variations can be used also for
simultaneous determination of magnetic field of a NS and distance
to the X-ray source (Popov 1999).

\begin{figure}
\epsfxsize=0.9\hsize
\centerline{\rotate[r]{\epsfbox{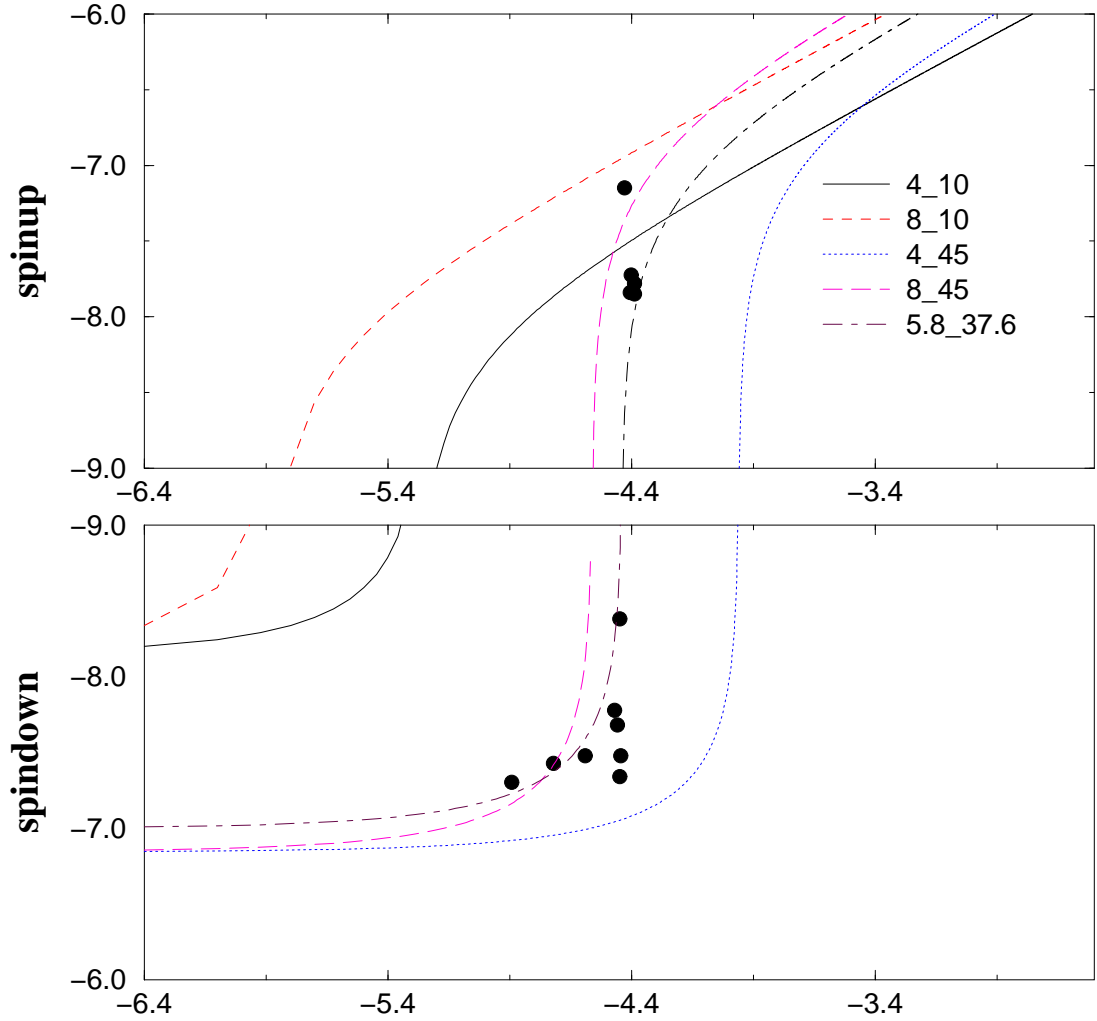}}}
\caption{Dependence of period derivative, $\dot p$, on the parameter $p^{7/3}f$, 
$f$-- observed flux, for GRO 1008-57. Both axis are in logarithmic scale.
Observations (Bildsten et al., 1997) are shown with black dots.
Five curves are plotted for disk accretion for different values of
distance to the pulsar and NS magnetic momentum. Solid curve: 
$d=4 \, {\rm kpc}$, $\mu=37.6\cdot 10^{30}\, {\rm G}\cdot {\rm cm}^3$. 
Dashed curve: $d=8 \, {\rm kpc}$, $\mu=37.6\cdot 10^{30}\, {\rm G}\cdot {\rm
cm}^3$.
Long dashed curve: $d=5.8 \, {\rm kpc}$, $\mu=10\cdot 10^{30}\, {\rm G}
\cdot {\rm cm}^3$.
Dot-dashed curve: $d=5.8 \, {\rm kpc}$, $\mu=45\cdot 10^{30}\, {\rm G}
\cdot {\rm cm}^3$. 
Dotted curve (the best fit): $d=5.8 \, {\rm kpc}$, 
$\mu=37.6\cdot 10^{30}\, {\rm G}\cdot
{\rm cm}^3$.}
\end{figure}

\begin{figure}
\epsfxsize=0.9\hsize
\centerline{\rotate[r]{\epsfbox{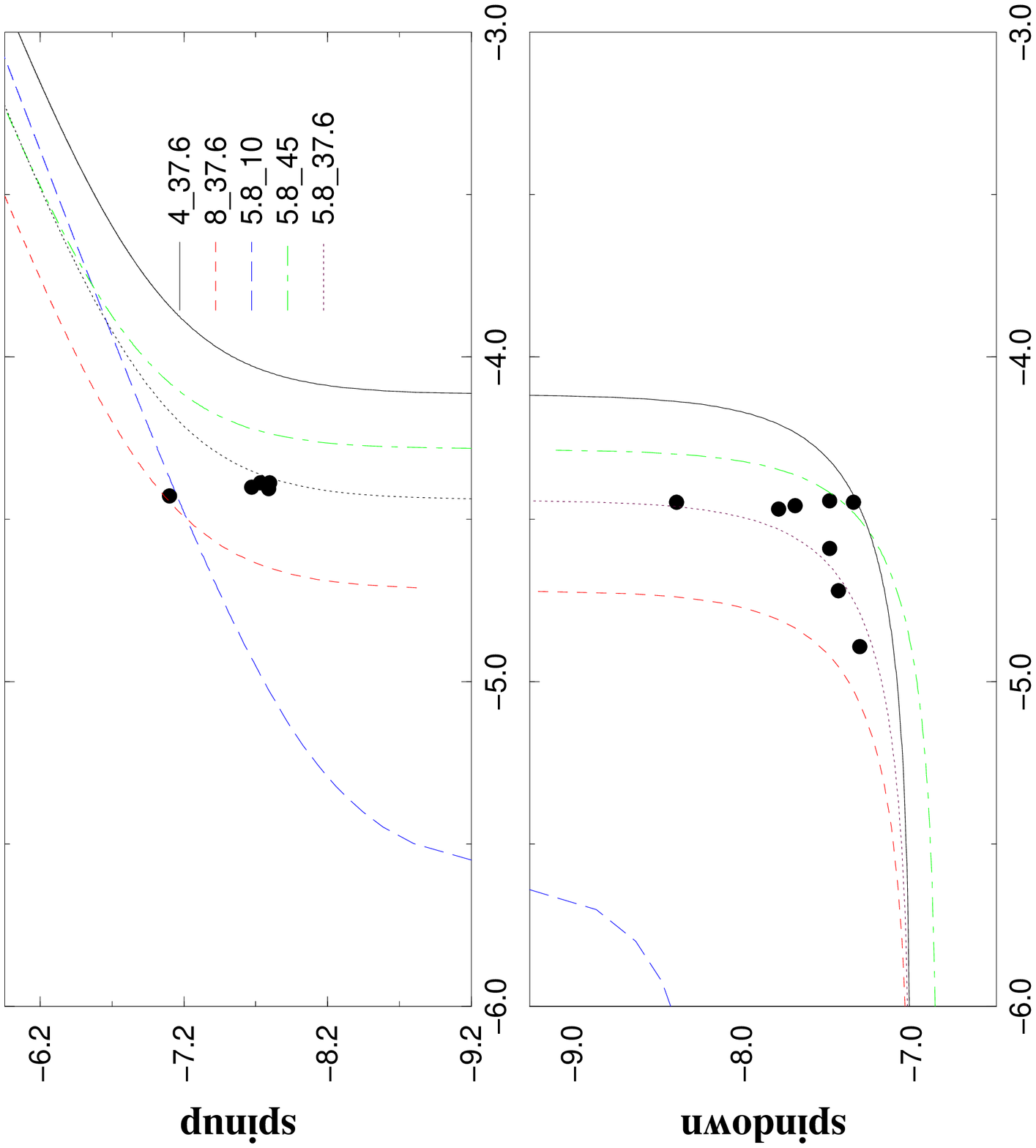}}}
\caption{Dependence of period derivative, $\dot p $, on the parameter $p^{7/3}f$, 
$f$-- observed flux, for GRO 1008-57. Both are axis in logarithmic scale.
Observations (Bildsten et al., 1997) are shown with black dots.
Five curves are plotted for disk accretion for different values of
distance to the pulsar and NS magnetic momentum. Solid curve: 
$d=4 \, {\rm kpc}$,
$\mu=10\cdot 10^{30}\, {\rm G}\cdot {\rm cm}^3$. 
Dashed curve: $d=8 \, {\rm kpc}$, 
$\mu=10\cdot 10^{30}\, {\rm G}\cdot {\rm cm}^3$.
Long dashed curve: $d=8 \, {\rm kpc}$, 
$\mu=45\cdot 10^{30}\, {\rm G}\cdot {\rm cm}^3$.
Dot-dashed curve (the best fit): $d=5.8 \, {\rm kpc}$, 
$\mu=37.6\cdot 10^{30}\,
{\rm G}\cdot {\rm cm}^3$. 
Dotted curve: $d=4 \, {\rm kpc}$, $\mu=45\cdot 10^{30}\, {\rm G}\cdot 
{\rm cm}^3$.}
\end{figure}

The method is based on several measurements
of period derivative, $\dot p$, and X-ray pulsar's flux, $f$.
Fitting distance, $d$, and magnetic momentum, $\mu$,
one can obtain good correspondence with the observed $p, \,\dot p$ and $f$,
and that way produce good estimates of distance and magnetic field
(see also another way of estimating of these parameters based
on the equilibrium period and spin-up measurements applied to GRO1744-28 in
(Joss \& Rappaport 1997) and (Rappaport \& Joss 1997)).

 Lets consider only disk accretion due to application of our method
to the system, in which most probably accretion is of the disk type. 
In that case one can write (see Lipunov 1982, 1992):

\begin{equation}
\dot p= \frac {4 \pi ^2 \mu ^2}{3\,G\,I\,M} - \sqrt{0.45}\, 2^{-1/14}\frac
{\mu^{2/7}}{I} \left(GM\right)^{-3/7} \left[p^{7/3}L\right]^{6/7}R^{6/7},
\end{equation}

where $L=4\pi d^2 \cdot f$ -- luminosity, $f$ -- the observed flux.

So, with some small uncertainty 
in the equation above we know all parameters ($I$, $M$, $R$ etc.) 
except $\mu$ and $d$.
Fitting observed points with them we can obtain estimates of $\mu$ and $d$.
Uncertainties mainly depend on applicability of that simple model.

 To illustrate the method, we apply it to the X-ray pulsar 
GRO J1008-57, discovered by BATSE (Bildsten et al., 1997). It is a 
$93.5 $ s X-ray pulsar, with the BATSE flux about $10^{-9}$ erg$\,$ cm$^{-2}$
s$^{-1}$. A 33 day outburst was observed by BATSE in August 1993.
The source was
identified with a Be-system 
with $\sim 135^d$ orbital period (Shrader et al. 1999).
We use here only 1993 outburst, described in Bildsten et al. (1997).

 Bildsten et al. (1997) show flux and frequency history
of the source with 1 day integration. In the maximum of the burst errors
are rather small, and we neglect them. Points with large errors were not used.

 We used standard values of NS parameters: $I=10^{45}\,$ g$\,$ cm$^2$, 
momentum of
inertia; $R=10\,$ km, NS radius; $M=1.4M_{\odot} $, NS mass.

 On figures 3-4 we show observations (as black dots)
and calculated curves (in the disk model,
see Shrader et al. (1999), who proposed a disk formation during the outbursts,
in contrast with Macomb et al. (1994), who proposed wind accretion)
on the plane $\dot p$ -- $p^{7/3} f$, where $f$ -- observed flux (logarithms
of these quantities are shown).
Curves were plotted for different values of the source distance, $d$,
and NS magnetic momentum, $\mu$. Spin-up and spin-down rates were obtained
from graphs in Bildsten et al. (1997).

 The best fit (both for spin-up and spin-down)
 gives $d\approx 5.8\, {\rm kpc}$ and $\mu\approx 37.6\cdot 10^{30}\,$ G
$\cdot$ cm$^3$.
It is shown on both figures. The distance is in correspondence with
the value in (Shrader et al. 1999), and such field value is not unusual
for NSs in general and for X-ray pulsars in particular (see, for example,
(Lipunov 1992) and (Bildsten et al. 1997)), and this value of $\mu$ is
consistent with maximum spin-down (see table 1). 
Tests on some other X-ray pulsars
with know distances and magnetic fields also showed good results.

 The method of distance and field estimates
is  approximate and depends on several assumptions (type of
accretion, specified values of $M, I, R$, etc.). 
Estimates of $\mu$, for example, can be only in rough correspondence with
determinations of magnetic field $B$ with hyrolines, 
if standard value of the NS radius,
$R=10$ km is used (see, for example,
the case of Her X-1 in (Lipunov 1992)). 
When the field and the distance are know with high precision
observations of period and flux observations can be used to put limits on
the equation of state (see e.g. Schaab \& Weigel 1999).

If one uses maximum spin-up, 
or maximum spin-down values to evaluate parameters of the pulsar,
then one can obtain values different from the best fit
(they are also shown on the figures): $d\approx 8 \,$ kpc, 
$\mu\approx 37.6\cdot 10^{30}\,$ G$\cdot$ cm$^3$
for maximum spin-up, and two values
for maximum
spin-down: $d\approx 4 \, {\rm kpc}$, 
$\mu\approx 37.6\cdot 10^{30}\,$ G$\cdot$ cm$^3$ and the one close to our
best fit (two similar values of maximum spin-down were observed
for different fluxes, but we mark, that formally maximum spin-down
corresponds to the values, which are close to our best fit). 
It can be used as an estimate of the errors of our method:
accuracy is about the factor of 2 in distance, and about the same value in
magnetic field, as can be seen from the figures.

Determination of magnetic field (and, probably, distance) only from X-ray
observations can be very useful in uncertain situations, for example, when
only X-ray observations
without precise localizations are available.\\

\noindent
{\bf Acknowledgments}

PSB thanks prof. Joss for discussions.

The work was supported by the RFBR (98-02-16801) and
the INTAS (96-0315) grants.


\begin{thebibliography}{}
\bibitem{} Bildsten, L. et al., 1997, ApJ Suppl. {\bf 113}, p. 367 
\bibitem{} Borkus, V.V., 1998, PhD dissertation, Space Research Institute,
Moscow
\bibitem{} Joss, P.C., \& Rappaport, S., 1997, in IAU Coll. 163 proc.,
Eds. D.T. Wickramasinghe et al., ASP Conference series, {\bf 121}, p.289
\bibitem{} Lipunov, V.M., 1992, ``Astrophysics of Neutron Stars'', 
Springer-Verlag
\bibitem{} Lipunov, V.M., 1982, AZh {\bf 59}, p. 888
\bibitem{} Macomb, D.J., Shrader, C.L., \& Schultz, A.B., 1994, ApJ
{\bf 437}, p. 845
\bibitem{}  Nagase, F., 1992, in
            Ginga Memorial Symposium (ISAS Symp. on Astroph.)
            eds. F.Makino \& F.Nagase, p.1
\bibitem{} Popov S.B., 1999, Astr. Astroph. Trans. (in press), 
{\it astro-ph/9906012}
\bibitem{} Rappaport, S, \& Joss, P.C., 1997, ApJ {\bf 486}, p. 435
\bibitem{} Schaab, C., Weigel, M.K., 1999, {\it astro-ph/9904211}
\bibitem{} Shrader, C.L., Sutaria, F.K., Singh, K.P., \& Macomb, D.J., 
1999, ApJ {\bf 512}, p. 920
\bibitem{} Wang, Y.-M., 1996, ApJ {\bf 465}, L111 
\end{thebibliography}
\end{document}